\documentclass[12pt]{iopart}
\usepackage{graphicx}
\usepackage{dcolumn}
\usepackage{bm}
\usepackage{comment}
\usepackage{color}
\usepackage{xr}
\usepackage[dvipsnames]{xcolor}
\usepackage{amssymb}
\usepackage{amsthm}
\usepackage{amsfonts}
\usepackage{listings}
\usepackage{enumerate}
\usepackage{latexsym}
\usepackage{psfrag}
\usepackage{bm}
\usepackage{blkarray}
\usepackage{array}
\usepackage[normalem]{ulem}
\usepackage{float}
\usepackage[colorlinks]{hyperref}
\usepackage{setstack}
\usepackage{array}

\def\g{\Gamma}
\newcolumntype{L}{>{$}l<{$}} 

\begin{document}

\title[Theoretical study of the topological properties of the ferromagnetic pyrite CoS$_2$]{Theoretical study of topological properties of ferromagnetic pyrite CoS$_2$}

\author{I. Robredo$^{1,2}$, N. B. M. Schröter$^{3,4}$, A. Reyes-Serrato$^{1,5}$, A. Bergara$^{1,2,6}$, F. de Juan$^{1,7}$, L. M. Schoop$^8$, M.G. Vergniory$^{1,9}$}

\address{$^1$ Donostia International Physics Center, 20018 Donostia-San Sebastian, Spain.}
\address{$^2$ Physics Department, University of the Basque Country UPV/EHU,
48080 Bilbao, Spain.}
\address{$^3$ Swiss Light Source, Paul Scherrer Institute, CH-5232 Villigen PSI, Switzerland.}
\address{$^4$Max Planck Institute of Microstructure Physics, 06120 Halle, Germany}
\address{$^5$ Centro de Nanociencias y Nanotecnología, Universidad Nacional
Autónoma de México, Apartado Postal 14, Ensenada Baja California,
22800, México.}
\address{$^6$ Centro de Física de Materiales, Centro Mixto CSIC -UPV/EHU, 20018 Donostia, Spain.}
\address{$^7$ IKERBASQUE, Basque Foundation for Science, Maria Diaz de Haro 3, 48013 Bilbao,
Spain}
\address{$^8$ Department of Chemistry, Princeton University, Princeton, NJ 08540, USA.}
\address{$^9$ Max Planck Institute for Chemical Physics of Solids , 01187 Dresden, Germany.}

\ead{inigo.robredo@dipc.org,maiagvergniory@dipc.org,maia.vergniory@cpfs.mpg.de}
\vspace{10pt}
\date{today}

\begin{abstract}

Since the discovery of the first topological material 15 years ago, the search for material realizations of novel topological phases has become the driving force of the field. 
While oftentimes we search for new materials, we forget that well established materials can also display very interesting topological properties. In this work, we revisit CoS${}_2$, a metallic ferromagnetic pyrite that has been extensively studied in the literature due to its magnetic properties. 
We study the topological features of its electronic band structure and identify Weyl nodes and Nodal lines, as well as a symmetry-protected 4-fold fermion close to the Fermi level. Looking at different surface cleavage planes, we observe both spin polarized Fermi arcs in the majority channel and drumhead states. These findings suggest that CoS$_2$ is a promising platform to study topological phenomena, as well as a good candidate for spintronic applications.
\end{abstract}

\vspace{2pc}

\section{Introduction}

Topological semimetals are a class of materials that display protected band crossings close to the Fermi level with nonzero Berry phase \cite{RevModPhys.88.035005,RevModPhys.90.015001,burkov_sm,felser_sm,vergnio,xu2020high}. In 3 dimensions (3D) one can find three types of topological band crossings, namely, nodes, lines and planes. When nodes have a linear dispersion in all directions and they carry a topological charge, they are named Weyl nodes. Weyl nodes can only exist in systems with broken inversion or time-reversal symmetry (TRS). They are stable even in the absence of symmetries and have very interesting surface states called Fermi arcs, that connect the surface projection of the nodes. From each Weyl node with topological charge 1, there will be a Fermi arc stemming from its projection in the surface. Important examples of inversion symmetry breaking Weyl semimetals are transition metal monophosphides \cite{TMmonophos}, MoTe${}_2$ \cite{Weyl_MoTe2} and TaAs \cite{PhysRevX.5.031013,Weyl_TaAs_1,Weyl_TaAs_2}. Contrary to the prolific discovery of inversion symmetry-broken Weyl semimetals, time reversal symmetry-broken (magnetic) ones are still scarce. They were first predicted in  pyrochlore iridates \cite{Weyl_pirochlore}, HgCr${}_2$Se${}_4$ \cite{HgCr2Se4} and the Heusler family of XCo$_{2}$Z (X=IVB or VB; Z=IVA or IIIA) \cite{PhysRevLett.117.236401}, and recently measured experimentally in EuCd$_2$As$_2$ \cite{PhysRevB.100.201102}, Co$_3$Sn$_2$S$_2$ \cite{Co3Sn2S2,Liu1282} and Co$_2$MnGa\cite{Belopolski1278}. These materials usually display a large anomalous Hall effect \cite{Co3Sn2S2}, which is important for both electronic and spintronic devices \cite{AHEapp}, and they played a fundamental role in the first experimental realization of the chiral anomaly \cite{PhysRev.177.2426}. Higher order generalizations of Weyl nodes have been proposed \cite{Bradlynaaf5037} and confirmed experimentally \cite{multi_exp,Hasan_CoSi,Hasan_CoSi_2,Takane_CoSi,PdSb2_paper}. These so called multifold fermions have higher Chern numbers, which produce more Fermi arcs in the surface. It has also been shown that space group chirality (no inversion or mirror symmetries) has an important effect on the connectivity of such Fermi arcs \cite{multi_exp,Exp_30}. Even if the search for multifold fermions has been focused on non-magnetic materials, there has been an increased shift of attention to magnetic systems. There are, however, few material predictions that can host magnetic multifold fermions \cite{Vergniory2018,doi:10.1126/sciadv.aar2317}, thus, making it a priority to find new material realizations of such multifold fermions.

Another notable topological feature are Nodal lines. These crossings are protected  by inversion-TRS, spin rotation or reflection symmetries \cite{Nodal_sym_1,Nodal_sym_2,Nodal_sym_3,Nodal_sym_4,Nodal_sym_5,Nodal_sym_6,Nodal_sym_7,Nodal_sym_8}. Compared to Weyl semimetals, Nodal line semimetals are more difficult to diagnose, since the topological invariants \cite{Nodal_topo} and, thus, the surface spectrum, are protected by crystalline symmetries, so surfaces not preserving them will not display topologically protected surface states \cite{Nodal_drum_1,Nodal_drum_2,Nodal_drum_3, PhysRevX.10.011026}. In the facets where symmetries are preserved, however, there will appear what are known as drumhead states, a set of surface states covering the projection of the nodal line in the surface. In the cases in which spin-orbit coupling (SOC) can be neglected, Nodal lines protected by inversion-TRS can occur at any point in the Brillouin zone (BZ), as well as in high symmetry planes. When including SOC, though, inversion-TRS can no longer protect the crossings, and these will generally be gapped. This is the case for spin rotation symmetry protected Nodal lines too, because SOC mixes spin components and gaps the crossings. Then, in the cases in which SOC cannot be neglected, only mirror symmetries can protect Nodal lines. This is why there are few examples of real materials that host Nodal lines when SOC is taken into account \cite{Nodal_soc_1,Nodal_drum_1}, and why finding new ones is exciting both theoretically and experimentally. Like in Weyl semimetals, Nodal line semimetals have recently been found to have large Spin Hall current \cite{Hou2021}, expanding the family of topological materials promising for topological electronic applications. Generalizing Nodal lines, non-symmorphic symmetries can force crossings on entire high symmetry planes in the boundary of the BZ. These are named nodal planes, which give rise to Topological Protectorates; regions of the Fermi surface that intersect the Nodal planes. They have been predicted to be large sources of Berry curvature \cite{Nodal_plane, Nodal_plane_2}.

Pyrites of the form XS${}_2$ (with X being Fe, Ni and Co) are a notable family of TRS-breaking materials that have been studied for decades due to the large size of pure crystals that can be found in nature or easily grown experimentally \cite{CoS2_grow_1,CoS2_mineral}. Among this family, CoS$_2$ has been experimentally confirmed to be ferromagnetic and extensively studied for its spin polarization and magnetic properties. Unlike many magnetic materials, CoS$_2$ is an itinerant ferromagnet \cite{itin_mag_1,itin_mag_2}, that is, the ferromagnetism does not stem from highly localized electrons. Thus, electron-electron interactions can be neglected when studying the electronic properties of the system. This implies that standard mean field approximations such as density functional theory (DFT) within the local spin density approximation (LSDA) are an excellent way to obtain reliable and accurate electronic properties \cite{PhysRevB.62.357}.

Topological condensed matter physics is a relatively new field of physics. Thus, it is possible that `old' materials could display topological properties that have been missed. This is the case on point; it has been recently reported experimentally that CoS$_2$ is a Weyl semimetal, with spin-polarized single and double Fermi arcs stemming from the projections of the Weyl nodes in the surface \cite{CoS2exp}. In this work, we perform an extensive study of the topological properties as a function of its crystal symmetries and also report several Nodal line structures near the Fermi level that survive the addition of SOC interaction. We also find novel drumhead surface states emerging from the Nodal lines.

The presence of both Nodal lines and Weyl nodes near the Fermi level, along with the big size of pure crystals and ferromagnetic nature of the pyrite, make CoS$_2$ a promising platform to probe topological phenomena in an accessible material that can be used in all sorts of electronic applications, from new magnetic memories to spin injector junctions, in which the spin polarized Fermi arcs will contribute to reduce the weight of the minority spin pocket \cite{Arc_inject_1}.

This work is structured as follows:  First we study the symmetry of the crystal and electronic structures. Next, we analyze the symmetry enforced 4-fold degeneracy arising at the $M=(\frac{1}{2},\frac{1}{2},0)$ point and the effect of spin-orbit coupling on it. Then, we study the Nodal lines and Weyl nodes in CoS${}_2$. Finally, we summarize the results of this work.

\section{Results and discussion}\label{crystal_structure}

\begin{figure}
    \centering
    \includegraphics[width=\linewidth]{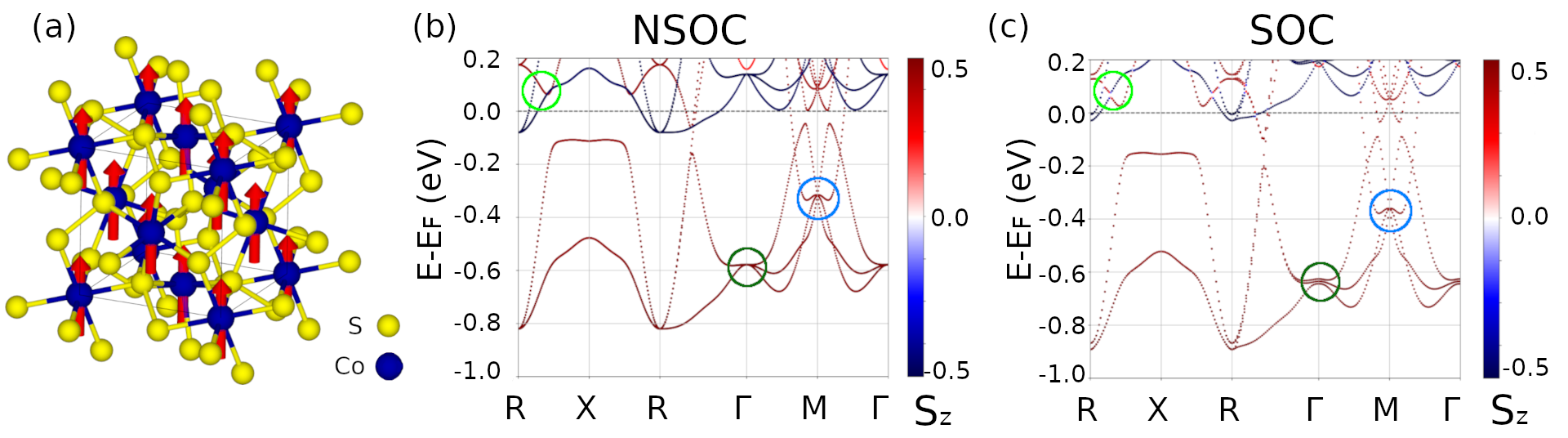}
    \caption{(a) Crystal and magnetic structure of CoS${}_2$. Electronic band structures near the Fermi level  without (b) and with (c) SOC interaction included. Light green, dark green and blue circles enclosed  gapless/gapped band crossing owing to SOC interaction mixing spins, 3-fold/2-fold degeneracy w/o and with SOC interaction included and the symmetry enforced 4-fold degeneracy, respectively.}
    \label{fig:electronicbands}
\end{figure}

CoS${}_2$ crystallizes in SG Pa$\bar 3$ (205). This is a non-symmorphic cubic Space Group, generated by \{I$|$0\}, \{C${}_{2z}|\frac{1}{2}0\frac{1}{2}$\} and \{C${}_{31}|$0\}, whose combination produces glide symmetry planes, $\{m_x|1/2,1/2,0\}$, $\{m_y|0,1/2,1/2\}$ and $\{m_z|1/2,0,1/2\}$. Cobalt atoms sit in the 4a Wyckoff position, while sulfur atoms sit in the 8c Wyckoff position. The structure is shown in Fig. \ref{fig:electronicbands}(a). Experimental measurements \cite{doi:10.1063/1.1656326, Teruya_2017} reveal that the system has ferromagnetic ordering, which mainly stems from the cobalt d orbitals. This is represented by red arrows on cobalt atoms in Fig. \ref{fig:electronicbands}(a). Experimental measurements determine that the magnetization is on the (100) direction family. We choose the direction of the magnetization in the z direction, without loss of generality.

When taking into account magnetization, the symmetry of the system is lowered. The 3-fold axes are broken, as well as TRS. The only surviving unitary symmetries are the 2-fold screw \{C${}_{2z}|\frac{1}{2}0\frac{1}{2}$\}, the glide $\{m_z|1/2,0,1/2\}$ and inversion symmetry. The screws in the x and y directions, as well as the glides orthogonal to $x$ and $y$ axis are preserved in combination with TRS. All these properties are summarized in its  magnetic space group, which is Pb'c'a (No. 61.436) \cite{BCS1,BCS2,BCS3}.

To study the electronic structure, we performed density functional theory (DFT) calculations as implemented in the Vienna ab initio simulation package (VASP) \cite{VASP1,VASP2,VASP3,VASP4}. As it was shown in previous studies \cite{CoS2exp, Brown_2005}, the appropriate  method to compute the electronic structure is the local spin density approximation (LSDA) with no Hubbard U correction \cite{PhysRevB.62.357}, with the Dudarev simplified exchange correlation term, together with PAW pseudopotentials. We used a grid of 7x7x7 k-points for the self-consistent calculation. 

The resulting band structures, with and without the effect of spin-orbit coupling are shown in Fig. \ref{fig:electronicbands}(b) and Fig. \ref{fig:electronicbands}(c). Notice that the bands below the Fermi level are polarized mainly in the direction of the majority spin (up, following the magnetization), with a small minority pocket at the $R$ point, as experimentally demonstrated in Ref. \cite{CoS2exp}.

Without SOC in a ferromagnet, the Schr\"odinger equation can be decoupled into spin up and spin down sectors where spin is labeled in the quantization axis of the ferromagnet. In one of the spin sectors, the symmetry thus appears to be the non-magnetic one Pa$\bar 3$ (205), regardless of the orientation of the magnetization. This artifice breaks down when SOC is included. Since SOC interaction has small effect on the band structure (see Fig. \ref{fig:electronicbands}(b) and Fig. \ref{fig:electronicbands}(c)), we expect the symmetry breaking effects to be small. In the band structure plots shown in Fig. \ref{fig:electronicbands}(b) and Fig. \ref{fig:electronicbands}(c), SOC couples both spins and opens gaps when spin-up and spin-down bands meet (see light green circles). In addition, the 3-fold degeneracy at the $\g$ point in the SOC-free plot is lifted when including SOC (see dark green circles). As expected, the gaps are small.

In Fig. \ref{fig:electronicbands}(b) we observe a 4-fold degeneracy near the Fermi level in k-point $M=(\frac{1}{2},\frac{1}{2},0)$ (see blue circles), which survives the addition of SOC as shown in Fig. \ref{fig:electronicbands}(c). Since we are interested in band crossings close to the Fermi level, we analyse this degeneracy. It is a symmetry-protected degeneracy, following the 4-dimensional irreducible co-representation at the $M$ point. Following Group Theory tools, we can construct the most general, symmetry allowed, low-energy Hamiltonian that describes the 4-fold degeneracy in the vicinity of the $M$ point. The resulting $\bf k\cdot p$ Hamiltonian at first order in momentum and magnetization\footnote{We include the magnetization as a finite parameter in the expansion, because we will be interested in computing the model on the other $M$ points later on. Then, we include a general magnetization $(B_x,B_y,B_z)$, even though only $B_z$ is non-zero in our case.} reads

\begin{equation}
    H(\boldsymbol{k})=B_x\boldsymbol{f\cdot\sigma}\otimes\tau_3+k_xv_xB_z\sigma_0\otimes\tau_1+ k_y\boldsymbol{v_y\cdot\sigma}\otimes\tau_1+k_zv_z\sigma_0\otimes\tau_3,
\end{equation}
where momentum $\boldsymbol{k}=(k_x,k_y,k_z)$ is measured from $M$, $\sigma,\,\tau$ are Pauli matrices, $B_x$ is the magnetic field in the x direction and $\boldsymbol{f}, v_x, \boldsymbol{v_y}, v_z$ are undetermined constants. This resembles the Hamiltonian of a Dirac fermion, but there are two main differences: first, the momentum-independent $B_x\boldsymbol{f\cdot\sigma}\otimes\tau_3$ term, which gaps the 4-fold when $B_x\neq0$ and second that both $k_x$ and $k_y$ go with $\tau_1$ and there is no $\tau_2$, so the resulting energy dispersion does not result in a 3D cone:

\begin{equation}
    E(\boldsymbol{k})=\pm\sqrt{(B_xf)^2+(v_xB_zk_x+v_yk_y)^2+(v_zk_z)^2}
\end{equation}
where $f=\sqrt{\boldsymbol{f}^2}$ and $v_y=\sqrt{\boldsymbol{v}_y^2}$. To see that this is not a cone, we can do a rotation of coordinates such that $k_1=v_xB_zk_x+v_yk_y$ and $k_2$ is orthogonal to $k_1$ and lies in the $k_z=0$ plane. In that case, the resulting energy dispersion would be cone-like in $k_1,\,k_z$ but completely flat in $k_2$. Thus, we cannot compute topological quantities such as the Chern number of half-filled bands, because they are completely degenerate in the $k_2$ direction.

Furthermore, we can rotate the model to find the corresponding $\bf k\cdot p$ model in the other M points, $M_2=(\frac{1}{2},0,\frac{1}{2})$ and $M_3=(0,\frac{1}{2},\frac{1}{2})$. $\{C_{31}|0\}$ symmetry transforms momentum and magnetization in the same way,  $C_{31}(k_x,k_y,k_z)=(k_z,k_x,k_y)$ and $C_{31}(B_x,B_y,B_z)=(B_z,B_x,B_y)$. Then,
\newpage
\begin{eqnarray}
   \hfill C_{31}M_1=C_{31}(\frac{1}{2},\frac{1}{2},0)=(0,\frac{1}{2},\frac{1}{2})=M_3 \nonumber\\
   \hfill C_{31}B_x=C_{31}(B,0,0)=(0,B,0)=B_y \nonumber\\
   \hfill C_{31}M_3=C_{31}(0,\frac{1}{2},\frac{1}{2})=(\frac{1}{2},0,\frac{1}{2})=M_2 \nonumber\\
   \hfill C_{31}B_y=C_{31}(0,B,0)=(0,0,B)=B_z.
\end{eqnarray}

\begin{figure} \label{fig:mpoints}
    \centering
    \includegraphics[width=\linewidth]{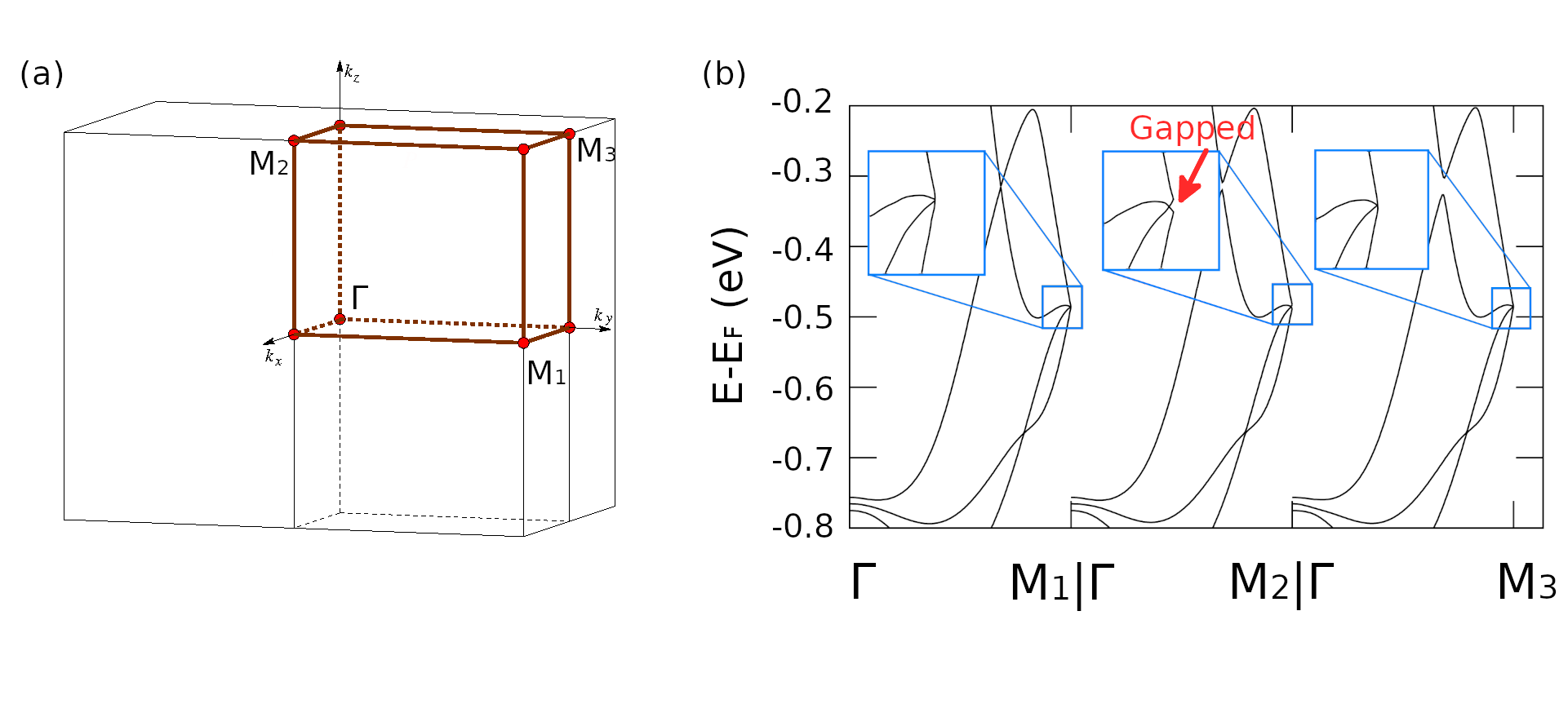}
    \caption{(a) Brillouin zone of the MSG Pb'c'a (No. 61.436) depicting the position of nonequivalent $M_i$ points (extracted from BCS \cite{BCS1,BCS2,BCS3}). (b) Dispersion from $\g$ to the different $M_i$ points. The red arrow indicates the  gap opening in $M_2$, as predicted by the $\bf k\cdot p$ model.}
\end{figure}

Notice that the momentum independent term that goes with $B_x$ in $M_1$ will go with $B_z$ in $M_2$. Since we have chosen the direction of the magnetization in the $z$ axis, we predict that the 4-fold will gap in the $M_2$ point. We confirmed this by computing the energy bands in the 3 $M_i$ points (see Fig. \ref{fig:mpoints}(a)) and we found two of them keep the band crossings ($M_1,M_3)$ and  the other one has a gap of 5.5 meV, as shown in Fig. \ref{fig:mpoints}(b). The size of the gap is directly related to the energy scale of the SOC interaction. Since SOC effect is small, the size of the gap must be small too.

Apart from 4-fold degeneracy at the $M_1$ and $M_3$ points, we also found that the planes $k_x=\pm0.5$ and $k_y=\pm0.5$ only have 2-dimensional magnetic irreducible co-representations. This implies that the bands will always be 2-fold degenerate at those planes, thus, they are Nodal planes. However, due to the presence of inversion symmetry, they will not have topological charge \cite{Nodal_plane}.

\begin{figure}
    \centering
    \includegraphics[width=1\linewidth]{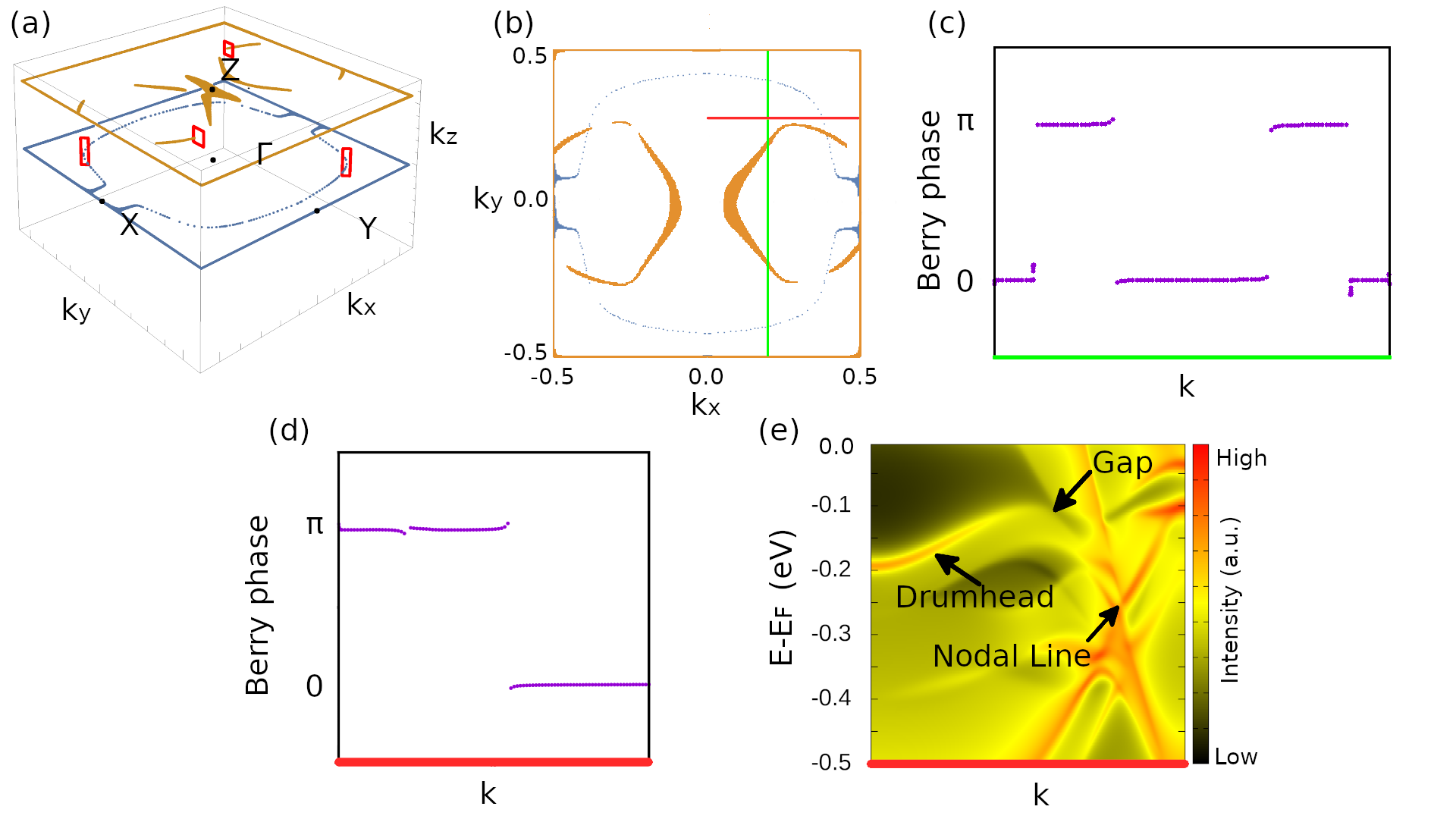}
    \caption{(a) Position of the Nodal lines in the BZ. Each red rectangle represents an integrated Berry phase of $\pi$. (b) Projection of the Nodal lines in the (001) direction. (c) and (d) Berry phase calculations on the paths depicted in (b). (e) Surface calculation in the (001) cleavage plane, computed along the red path shown in (b). Notice that the drumhead surface states gap in regions where the Berry phase is equal to $0$.}
    \label{fig:soc_67}
\end{figure}
\begin{figure}
    \centering
    \includegraphics[width=\linewidth]{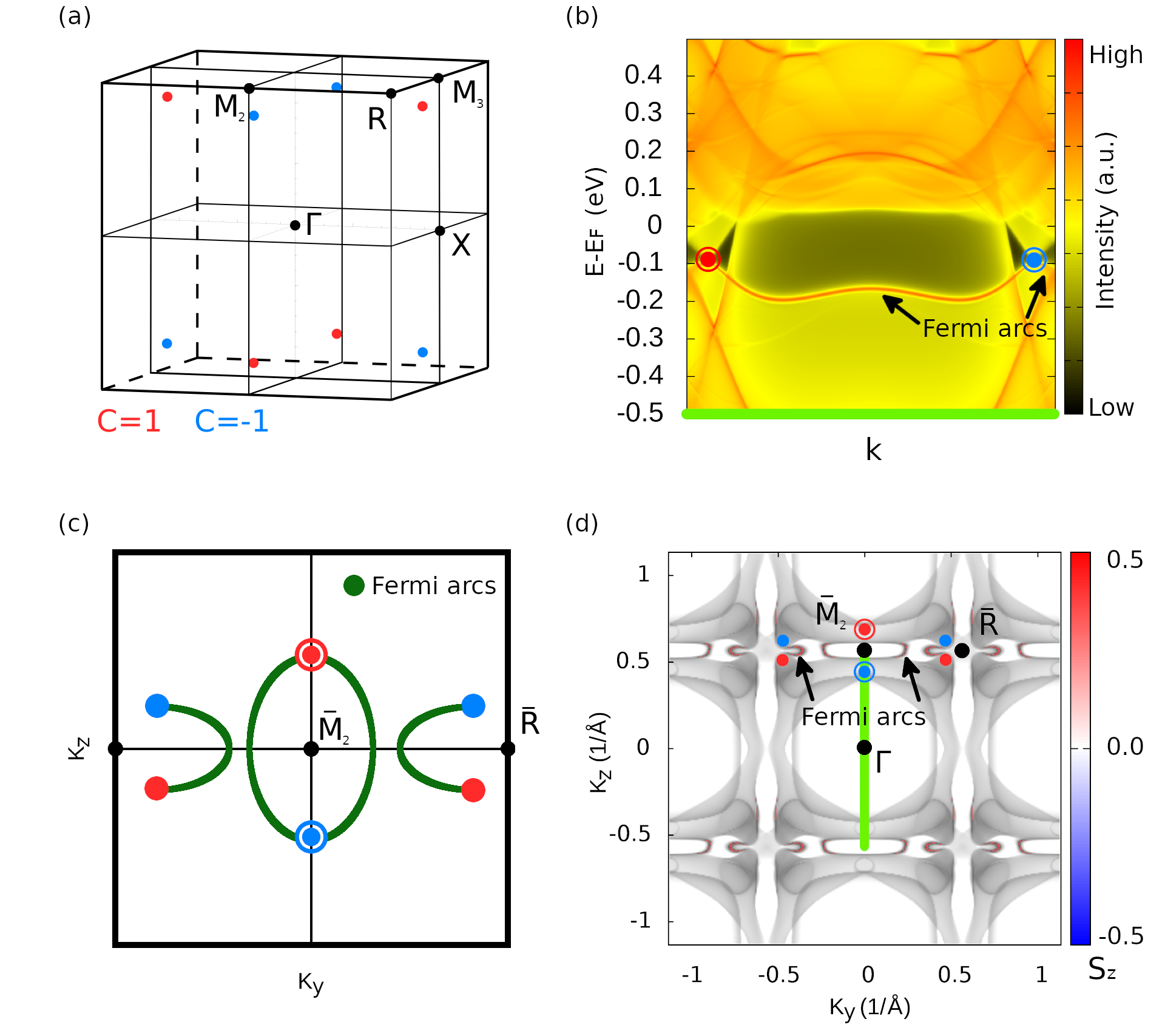}
    \caption{(a) Position of the Weyl nodes in the BZ.  (b) Surface calculation in the (100) cleavage plane in the path depicted in (d). (c) Zoomed-in schematic depiction of the surface states in the (100) cleavage plane near the $\bar M_2$-$\bar R$ line. The location of the projection of the Weyls into the surface is shown by red and blue dots. Dots with a surrounding circle represent the projection of two Weyl nodes with equal chirality. (d) Spin polarization of Fermi arcs in the surface calculation. They are polarized in the direction of the majority spin, just like the bulk bands.}
    \label{fig:soc-surface}
\end{figure}

We now study the topological properties of CoS${}_2$ focusing on both Weyl nodes and Nodal lines. The calculations in what follows are based on the interpolated Tight Binding model in the Wannier basis \cite{Wannier90} constructed from VASP ab-initio calculation \cite{VASP1,VASP2,VASP3,VASP4}. We performed Berry phase calculations, as well as surface calculations as implemented in WannierTools \cite{WannierTools}.

We focus on the crossings between the last two valence bands near the Fermi level (the two top bands polarized in the majority spin in Fig. \ref{fig:electronicbands}(c)). We found two nodal lines, one in the plane $k_z=0$ and another one in $k_z=\pi$. Both planes are left invariant by $\{m_z|1/2,0,1/2\}$ (see Fig. \ref{fig:soc_67}(a)), which is the symmetry that protects both Nodal lines. In order to characterize the topological protection of the Nodal line, we computed the Berry phase around a loop enclosing it. The location of the Berry phase integration paths (red rectangles) is shown Fig. \ref{fig:soc_67}(a). Each red rectangle represents an integrated Berry phase of $\pi$, thus, non-trivial $Z_2$ index. We conclude that the Nodal lines in both planes are topologically protected.

In Fig. \ref{fig:soc_67}(b) we show the projection of both Nodal lines in the (001) direction. We showed that the Nodal lines are topologically protected, however, which region of the $k_z=0$ plane will host surface states depends on the choice of surface termination \cite{Naumov_origin}. We fix our unit cell as shown in Fig. \ref{fig:electronicbands}(a) and compute the Berry phases (integrating on $k_z$) along the momentum paths depicted in Fig. \ref{fig:soc_67}(b). The results of the calculation are shown in Fig. \ref{fig:soc_67}(c) and Fig. \ref{fig:soc_67}(d). The calculations show that the Berry phase inside the blue Nodal line is equal to $\pi$, while the Berry phase outside, or inside the projection of both Nodal lines is equal to 0. Thus, we will only find drumhead states in the region where nodal lines do not overlap and the Berry phase is equal to $\pi$. Berry phase and surface spectrum calculations shown in Fig \ref{fig:soc_67} (d) and Fig. \ref{fig:soc_67}(e) respectively, have been performed in the same momentum path, as depicted in Fig. \ref{fig:soc_67}(b). We choose the surface termination to coincide with the periodic repetition of the unit cell. We observe that drumhead surface states survive only in regions where the Berry phase of Fig. \ref{fig:soc_67}(d) is equal to $\pi$, while they gap in regions where it is equal to 0.

By carefully examining the region close to the Fermi level we found a total of 8 Weyl points close to the Fermi level, whose location in momentum space is displayed in Fig. \ref{fig:soc-surface}(a). In Fig. \ref{fig:soc-surface}(b), we show (100) surface spectrum for $k_y=0$, along the momentum path depicted in Fig. \ref{fig:soc-surface}(d) in green. In this path, we cross the projection of 4 Weyl nodes, projected pairwise with the same chirality. We can see two bright surface states that connect the projection of the Weyl nodes, one within the BZ and the other going through the boundary. Notice that Fermi arcs connect Weyl nodes across the boundary of the BZ but they avoid the Nodal planes, which are located at $k_y=\pm\pi$. In the (100) cleavage plane, we can observe 4 different Fermi arcs as depicted in Fig. \ref{fig:soc-surface}(c). The Fermi arc close to the $\bar R$ point connects the projection of two opposite chirality Weyl nodes. Near the $\bar M$ points, though, we can see two Fermi arcs connecting the projections, instead of one. This is due to pairs of Weyl nodes (with the same chirality) being projected in the same point. In Fig. \ref{fig:soc-surface}(d), we see the Fermi surface calculation on the (100) surface, schematized in Fig. \ref{fig:soc-surface}(c). The connection of the Fermi arcs is the same one as in the scheme of Fig. \ref{fig:soc-surface}(c), with single Weyl nodes connected by a single Fermi arc and double Weyl nodes connected by two Fermi arcs. From the ab initio calculation we know that the bands in the relevant energy window are predominantly polarized  in the direction of the majority spin. We computed the spin polarization of the Fermi arcs to see if they inherit the polarization from the bulk. We show in Fig \ref{fig:soc-surface} (d) that the Fermi arcs are completely polarized in the direction of the majority spin, as is the case in the bulk.

Our results have implications in a wide range of fields. In Quantum Optics, it has been shown that Weyl semimetals could be used to realize a Veselago lens for electrons \cite{Weyl_app_1} (negative refractive index). It has also been confirmed experimentally that Weyl semimetals have a quantized circular photogalvanic effect \cite{Weyl_app_2, Weyl_app_4} due to their topological charge. In the field of electronics, Weyl semimetals show signatures of the chiral anomaly \cite{GdPtBi}. There are materials that show a large magnetorresistance, such as the family of TaAs \cite{PhysRevX.5.031013}, which can be harnessed for the next generation of memory devices. The spin polarization of the Fermi arcs can also influence the performance of a Weyl semimetal as a spin injector, improving the expected bulk results \cite{Weyl_app_3}.

\section{Conclusions}
In this work we revisited the recently discovered Weyl semimetal CoS${}_2$ ferromagnetic pyrite. We analyzed both Nodal lines and Weyl nodes found on the ferromagnetic system. We found two nodal lines close to the Fermi level, with surface drumhead states in the (001) surface. We also found a family of Weyl nodes close to the Fermi level, with their distinctive Fermi arcs on the (100) surface. These surface states increase the number of available electrons in the surface, thus improving its metallic properties for junctions. Following the spin polarization in the direction of the majority in the bulk, we checked if the Fermi arcs are polarized in the same direction too. We answer this question in the affirmative, adding this material to the family of spin polarized Fermi arcs in real materials \cite{Spin_arc_1}. The Fermi arcs would contribute to the spin majority polarization, making it an interesting material for spintronics, like a possible spin injector \cite{Spin_arc_2}.
 
\section{Acknowledgements}

M.G.V. and I.R. acknowledge the Spanish Ministerio de Ciencia e Innovacion (grant number PID2019109905GB-C21) and Programa Red Guipuzcoana de Ciencia, Tecnología e Innovación 2021 No. 2021-CIEN-000070-01 Gipuzkoa Next. A.R.S. is grateful to PASPA-DGAPA-UNAM for the sabbatical scholarship and to the DIPC for their support during the research stay. A.B. acknowledges financial support from the Spanish Ministry of Science and Innovation (PID2019-105488GBI00). N.B.M.S. was supported by Microsoft. This work was supported by the Gordon and Betty Moore Foundation through Grant GBMF9064 to L.M.S.

\section*{References}
\bibliographystyle{iopart-num}
\bibliography{biblio.bib}

\providecommand{\newblock}{}
\begin{thebibliography}{10}
\expandafter\ifx\csname url\endcsname\relax
  \def\url#1{{\tt #1}}\fi
\expandafter\ifx\csname urlprefix\endcsname\relax\def\urlprefix{URL }\fi
\providecommand{\eprint}[2][]{\url{#2}}

\bibitem{RevModPhys.88.035005}
Chiu C~K, Teo J~C~Y, Schnyder A~P and Ryu S 2016 {\em Rev. Mod. Phys.\/} {\bf
  88}(3) 035005
  \urlprefix\url{https://link.aps.org/doi/10.1103/RevModPhys.88.035005}

\bibitem{RevModPhys.90.015001}
Armitage N~P, Mele E~J and Vishwanath A 2018 {\em Rev. Mod. Phys.\/} {\bf
  90}(1) 015001
  \urlprefix\url{https://link.aps.org/doi/10.1103/RevModPhys.90.015001}

\bibitem{burkov_sm}
Burkov A 2018 {\em Annual Review of Condensed Matter Physics\/} {\bf 9}
  359--378 (\textit{Preprint}
  \eprint{https://doi.org/10.1146/annurev-conmatphys-033117-054129})
  \urlprefix\url{https://doi.org/10.1146/annurev-conmatphys-033117-054129}

\bibitem{felser_sm}
Yan B and Felser C 2017 {\em Annual Review of Condensed Matter Physics\/} {\bf
  8} 337--354 (\textit{Preprint}
  \eprint{https://doi.org/10.1146/annurev-conmatphys-031016-025458})
  \urlprefix\url{https://doi.org/10.1146/annurev-conmatphys-031016-025458}

\bibitem{vergnio}
Vergniory M~G, Elcoro L, Felser C, Regnault N, Bernevig B~A and Wang Z 2019
  {\em Nature\/} {\bf 566} 480--485

\bibitem{xu2020high}
Xu Y, Elcoro L, Song Z~D, Wieder B~J, Vergniory M, Regnault N, Chen Y, Felser C
  and Bernevig B~A 2020 {\em Nature\/} {\bf 586} 702--707

\bibitem{TMmonophos}
Weng H, Fang C, Fang Z, Bernevig B~A and Dai X 2015 {\em Phys. Rev. X\/} {\bf
  5}(1) 011029
  \urlprefix\url{https://link.aps.org/doi/10.1103/PhysRevX.5.011029}

\bibitem{Weyl_MoTe2}
Sun Y, Wu S~C, Ali M~N, Felser C and Yan B 2015 {\em Phys. Rev. B\/} {\bf
  92}(16) 161107
  \urlprefix\url{https://link.aps.org/doi/10.1103/PhysRevB.92.161107}

\bibitem{PhysRevX.5.031013}
Lv B~Q, Weng H~M, Fu B~B, Wang X~P, Miao H, Ma J, Richard P, Huang X~C, Zhao
  L~X, Chen G~F, Fang Z, Dai X, Qian T and Ding H 2015 {\em Phys. Rev. X\/}
  {\bf 5}(3) 031013
  \urlprefix\url{https://link.aps.org/doi/10.1103/PhysRevX.5.031013}

\bibitem{Weyl_TaAs_1}
Yang L~X, Liu Z~K, Sun Y, Peng H, Yang H~F, Zhang T, Zhou B, Zhang Y, Guo Y~F,
  Rahn M, Prabhakaran D, Hussain Z, Mo S~K, Felser C, Yan B and Chen Y~L 2015
  {\em Nature Physics\/} {\bf 11} 728--732 ISSN 1745-2481
  \urlprefix\url{https://doi.org/10.1038/nphys3425}

\bibitem{Weyl_TaAs_2}
Xu S~Y, Belopolski I, Alidoust N, Neupane M, Bian G, Zhang C, Sankar R, Chang
  G, Yuan Z, Lee C~C, Huang S~M, Zheng H, Ma J, Sanchez D~S, Wang B, Bansil A,
  Chou F, Shibayev P~P, Lin H, Jia S and Hasan M~Z 2015 {\em Science\/} {\bf
  349} 613--617 ISSN 0036-8075 (\textit{Preprint}
  \eprint{https://science.sciencemag.org/content/349/6248/613.full.pdf})
  \urlprefix\url{https://science.sciencemag.org/content/349/6248/613}

\bibitem{Weyl_pirochlore}
Wan X, Turner A~M, Vishwanath A and Savrasov S~Y 2011 {\em Phys. Rev. B\/} {\bf
  83}(20) 205101
  \urlprefix\url{https://link.aps.org/doi/10.1103/PhysRevB.83.205101}

\bibitem{HgCr2Se4}
Xu G, Weng H, Wang Z, Dai X and Fang Z 2011 {\em Phys. Rev. Lett.\/} {\bf
  107}(18) 186806
  \urlprefix\url{https://link.aps.org/doi/10.1103/PhysRevLett.107.186806}

\bibitem{PhysRevLett.117.236401}
Wang Z, Vergniory M~G, Kushwaha S, Hirschberger M, Chulkov E~V, Ernst A, Ong
  N~P, Cava R~J and Bernevig B~A 2016 {\em Phys. Rev. Lett.\/} {\bf 117}(23)
  236401
  \urlprefix\url{https://link.aps.org/doi/10.1103/PhysRevLett.117.236401}

\bibitem{PhysRevB.100.201102}
Soh J~R, de~Juan F, Vergniory M~G, Schr\"oter N~B~M, Rahn M~C, Yan D~Y, Jiang
  J, Bristow M, Reiss P, Blandy J~N, Guo Y~F, Shi Y~G, Kim T~K, McCollam A,
  Simon S~H, Chen Y, Coldea A~I and Boothroyd A~T 2019 {\em Phys. Rev. B\/}
  {\bf 100}(20) 201102
  \urlprefix\url{https://link.aps.org/doi/10.1103/PhysRevB.100.201102}

\bibitem{Co3Sn2S2}
Morali N, Batabyal R, Nag P~K, Liu E, Xu Q, Sun Y, Yan B, Felser C, Avraham N
  and Beidenkopf H 2019 {\em Science\/} {\bf 365} 1286--1291 ISSN 0036-8075
  (\textit{Preprint}
  \eprint{https://science.sciencemag.org/content/365/6459/1286.full.pdf})
  \urlprefix\url{https://science.sciencemag.org/content/365/6459/1286}

\bibitem{Liu1282}
Liu D~F, Liang A~J, Liu E~K, Xu Q~N, Li Y~W, Chen C, Pei D, Shi W~J, Mo S~K,
  Dudin P, Kim T, Cacho C, Li G, Sun Y, Yang L~X, Liu Z~K, Parkin S~S~P, Felser
  C and Chen Y~L 2019 {\em Science\/} {\bf 365} 1282--1285 ISSN 0036-8075
  (\textit{Preprint}
  \eprint{https://science.sciencemag.org/content/365/6459/1282.full.pdf})
  \urlprefix\url{https://science.sciencemag.org/content/365/6459/1282}

\bibitem{Belopolski1278}
Belopolski I, Manna K, Sanchez D~S, Chang G, Ernst B, Yin J, Zhang S~S, Cochran
  T, Shumiya N, Zheng H, Singh B, Bian G, Multer D, Litskevich M, Zhou X, Huang
  S~M, Wang B, Chang T~R, Xu S~Y, Bansil A, Felser C, Lin H and Hasan M~Z 2019
  {\em Science\/} {\bf 365} 1278--1281 ISSN 0036-8075 (\textit{Preprint}
  \eprint{https://science.sciencemag.org/content/365/6459/1278.full.pdf})
  \urlprefix\url{https://science.sciencemag.org/content/365/6459/1278}

\bibitem{AHEapp}
Liu C~X, Zhang S~C and Qi X~L 2016 {\em Annual Review of Condensed Matter
  Physics\/} {\bf 7} 301--321 (\textit{Preprint}
  \eprint{https://doi.org/10.1146/annurev-conmatphys-031115-011417})
  \urlprefix\url{https://doi.org/10.1146/annurev-conmatphys-031115-011417}

\bibitem{PhysRev.177.2426}
Adler S~L 1969 {\em Phys. Rev.\/} {\bf 177}(5) 2426--2438
  \urlprefix\url{https://link.aps.org/doi/10.1103/PhysRev.177.2426}

\bibitem{Bradlynaaf5037}
Bradlyn B, Cano J, Wang Z, Vergniory M~G, Felser C, Cava R~J and Bernevig B~A
  2016 {\em Science\/} {\bf 353} ISSN 0036-8075 (\textit{Preprint}
  \eprint{http://science.sciencemag.org/content/353/6299/aaf5037.full.pdf})
  \urlprefix\url{http://science.sciencemag.org/content/353/6299/aaf5037}

\bibitem{multi_exp}
Schr{\"o}ter N~B~M, Pei D, Vergniory M~G, Sun Y, Manna K, de~Juan F, Krieger
  J~A, S{\"u}ss V, Schmidt M, Dudin P, Bradlyn B, Kim T~K, Schmitt T, Cacho C,
  Felser C, Strocov V~N and Chen Y 2019 {\em Nature Physics\/} {\bf 15}
  759--765 ISSN 1745-2481
  \urlprefix\url{https://doi.org/10.1038/s41567-019-0511-y}

\bibitem{Hasan_CoSi}
Xu X, Wang X, Cochran T~A, Sanchez D~S, Chang G, Belopolski I, Wang G, Liu Y,
  Tien H~J, Gui X, Xie W, Hasan M~Z, Chang T~R and Jia S 2019 {\em Phys. Rev.
  B\/} {\bf 100}(4) 045104
  \urlprefix\url{https://link.aps.org/doi/10.1103/PhysRevB.100.045104}

\bibitem{Hasan_CoSi_2}
Sanchez D~S, Belopolski I, Cochran T~A, Xu X, Yin J~X, Chang G, Xie W, Manna K,
  S{\"u}{\ss} V, Huang C~Y, Alidoust N, Multer D, Zhang S~S, Shumiya N, Wang X,
  Wang G~Q, Chang T~R, Felser C, Xu S~Y, Jia S, Lin H and Hasan M~Z 2019 {\em
  Nature\/} {\bf 567} 500--505 ISSN 1476-4687
  \urlprefix\url{https://doi.org/10.1038/s41586-019-1037-2}

\bibitem{Takane_CoSi}
Takane D, Wang Z, Souma S, Nakayama K, Nakamura T, Oinuma H, Nakata Y, Iwasawa
  H, Cacho C, Kim T, Horiba K, Kumigashira H, Takahashi T, Ando Y and Sato T
  2019 {\em Phys. Rev. Lett.\/} {\bf 122}(7) 076402
  \urlprefix\url{https://link.aps.org/doi/10.1103/PhysRevLett.122.076402}

\bibitem{PdSb2_paper}
Kumar N, Yao M, Nayak J, Vergniory M~G, Bannies J, Wang Z, Schröter N~B~M,
  Strocov V~N, Müchler L, Shi W, Rienks E~D~L, Mañes J~L, Shekhar C, Parkin
  S~S~P, Fink J, Fecher G~H, Sun Y, Bernevig B~A and Felser C 2020 {\em
  Advanced Materials\/} {\bf 32} 1906046 (\textit{Preprint}
  \eprint{https://onlinelibrary.wiley.com/doi/pdf/10.1002/adma.201906046})
  \urlprefix\url{https://onlinelibrary.wiley.com/doi/abs/10.1002/adma.201906046}

\bibitem{Exp_30}
Schröter N~B~M, Stolz S, Manna K, de~Juan F, Vergniory M~G, Krieger J~A, Pei
  D, Schmitt T, Dudin P, Kim T~K and et~al 2020 {\em Science\/} {\bf 369}
  179–183 ISSN 1095-9203
  \urlprefix\url{http://dx.doi.org/10.1126/science.aaz3480}

\bibitem{Vergniory2018}
Vergniory M~G, Elcoro L, Orlandi F, Balke B, Chan Y~H, Nuss J, Schnyder A~P and
  Schoop L~M 2018 {\em The European Physical Journal B\/} {\bf 91} 213 ISSN
  1434-6036 \urlprefix\url{https://doi.org/10.1140/epjb/e2018-90302-7}

\bibitem{doi:10.1126/sciadv.aar2317}
Schoop L~M, Topp A, Lippmann J, Orlandi F, Müchler L, Vergniory M~G, Sun Y,
  Rost A~W, Duppel V, Krivenkov M, Sheoran S, Manuel P, Varykhalov A, Yan B,
  Kremer R~K, Ast C~R and Lotsch B~V 2018 {\em Science Advances\/} {\bf 4}
  eaar2317

\bibitem{Nodal_sym_1}
Chiu C~K and Schnyder A~P 2014 {\em Phys. Rev. B\/} {\bf 90}(20) 205136
  \urlprefix\url{https://link.aps.org/doi/10.1103/PhysRevB.90.205136}

\bibitem{Nodal_sym_2}
Yang B~J, Bojesen T~A, Morimoto T and Furusaki A 2017 {\em Phys. Rev. B\/} {\bf
  95}(7) 075135
  \urlprefix\url{https://link.aps.org/doi/10.1103/PhysRevB.95.075135}

\bibitem{Nodal_sym_3}
{Bian} G, {Chang} T~R, {Sankar} R, {Xu} S~Y, {Zheng} H, {Neupert} T, {Chiu}
  C~K, {Huang} S~M, {Chang} G, {Belopolski} I, {Sanchez} D~S, {Neupane} M,
  {Alidoust} N, {Liu} C, {Wang} B, {Lee} C~C, {Jeng} H~T, {Bansil} A, {Chou} F,
  {Lin} H and {Zahid Hasan} M 2015 {\em arXiv e-prints\/} arXiv:1505.03069
  (\textit{Preprint} \eprint{1505.03069})

\bibitem{Nodal_sym_4}
Yamakage A, Yamakawa Y, Tanaka Y and Okamoto Y 2016 {\em Journal of the
  Physical Society of Japan\/} {\bf 85} 013708 (\textit{Preprint}
  \eprint{https://doi.org/10.7566/JPSJ.85.013708})
  \urlprefix\url{https://doi.org/10.7566/JPSJ.85.013708}

\bibitem{Nodal_sym_5}
Xie L~S, Schoop L~M, Seibel E~M, Gibson Q~D, Xie W and Cava R~J 2015 {\em APL
  Materials\/} {\bf 3} 083602 (\textit{Preprint}
  \eprint{https://doi.org/10.1063/1.4926545})
  \urlprefix\url{https://doi.org/10.1063/1.4926545}

\bibitem{Nodal_sym_6}
Kim Y, Wieder B~J, Kane C~L and Rappe A~M 2015 {\em Phys. Rev. Lett.\/} {\bf
  115}(3) 036806
  \urlprefix\url{https://link.aps.org/doi/10.1103/PhysRevLett.115.036806}

\bibitem{Nodal_sym_7}
{Zeng} M, {Fang} C, {Chang} G, {Chen} Y~A, {Hsieh} T, {Bansil} A, {Lin} H and
  {Fu} L 2015 {\em arXiv e-prints\/} arXiv:1504.03492 (\textit{Preprint}
  \eprint{1504.03492})

\bibitem{Nodal_sym_8}
Huang H, Liu J, Vanderbilt D and Duan W 2016 {\em Phys. Rev. B\/} {\bf 93}(20)
  201114 \urlprefix\url{https://link.aps.org/doi/10.1103/PhysRevB.93.201114}

\bibitem{Nodal_topo}
Fang C, Chen Y, Kee H~Y and Fu L 2015 {\em Phys. Rev. B\/} {\bf 92}(8) 081201
  \urlprefix\url{https://link.aps.org/doi/10.1103/PhysRevB.92.081201}

\bibitem{Nodal_drum_1}
{Bian} G, {Chang} T~R, {Zheng} H, {Velury} S, {Xu} S~Y, {Neupert} T, {Chiu}
  C~K, {Huang} S~M, {Sanchez} D~S, {Belopolski} I, {Alidoust} N, {Chen} P~J,
  {Chang} G, {Bansil} A, {Jeng} H~T, {Lin} H and {Hasan} M~Z 2016 {\em Phys Rev
  B\/} {\bf 93} 121113 (\textit{Preprint} \eprint{1508.07521})

\bibitem{Nodal_drum_2}
Weng H, Liang Y, Xu Q, Yu R, Fang Z, Dai X and Kawazoe Y 2015 {\em Phys. Rev.
  B\/} {\bf 92}(4) 045108
  \urlprefix\url{https://link.aps.org/doi/10.1103/PhysRevB.92.045108}

\bibitem{Nodal_drum_3}
Yu R, Weng H, Fang Z, Dai X and Hu X 2015 {\em Phys. Rev. Lett.\/} {\bf 115}(3)
  036807
  \urlprefix\url{https://link.aps.org/doi/10.1103/PhysRevLett.115.036807}

\bibitem{PhysRevX.10.011026}
Muechler L, Topp A, Queiroz R, Krivenkov M, Varykhalov A, Cano J, Ast C~R and
  Schoop L~M 2020 {\em Phys. Rev. X\/} {\bf 10}(1) 011026
  \urlprefix\url{https://link.aps.org/doi/10.1103/PhysRevX.10.011026}

\bibitem{Nodal_soc_1}
Bian G, Chang T~R, Sankar R, Xu S~Y, Zheng H, Neupert T, Chiu C~K, Huang S~M,
  Chang G, Belopolski I, Sanchez D~S, Neupane M, Alidoust N, Liu C, Wang B, Lee
  C~C, Jeng H~T, Zhang C, Yuan Z, Jia S, Bansil A, Chou F, Lin H and Hasan M~Z
  2016 {\em Nature Communications\/} {\bf 7} 10556 ISSN 2041-1723
  \urlprefix\url{https://doi.org/10.1038/ncomms10556}

\bibitem{Hou2021}
Hou W, Liu J, Zuo X, Xu J, Zhang X, Liu D, Zhao M, Zhu Z~G, Luo H~G and Zhao W
  2021 {\em npj Computational Materials\/} {\bf 7} 37 ISSN 2057-3960
  \urlprefix\url{https://doi.org/10.1038/s41524-021-00504-w}

\bibitem{Nodal_plane}
Wilde M~A, Dodenh{\"o}ft M, Niedermayr A, Bauer A, Hirschmann M~M, Alpin K,
  Schnyder A~P and Pfleiderer C 2021 {\em Nature\/} {\bf 594} 374--379 ISSN
  1476-4687 \urlprefix\url{https://doi.org/10.1038/s41586-021-03543-x}

\bibitem{Nodal_plane_2}
Xiao M, Ye L, Qiu C, He H, Liu Z and Fan S 2020 {\em Science Advances\/} {\bf
  6} (\textit{Preprint}
  \eprint{https://advances.sciencemag.org/content/6/8/eaav2360.full.pdf})
  \urlprefix\url{https://advances.sciencemag.org/content/6/8/eaav2360}

\bibitem{CoS2_grow_1}
Wu P~F, Shi j~B, Cheng B, Lin H and Lee H 2019 {\em Materials Science Forum\/}
  {\bf 947} 91--95

\bibitem{CoS2_mineral}
Paul F 1945 {\em American Mineralogist\/} {\bf 30} 483--497

\bibitem{itin_mag_1}
Matthias B, Clogston A, Williams H, Corenzwit E and Sherwood R 1961 {\em
  Physical Review Letters\/} {\bf 7} 7

\bibitem{itin_mag_2}
Matthias B and Bozorth R 1958 {\em Physical Review\/} {\bf 109} 604

\bibitem{PhysRevB.62.357}
Kwon S~K, Youn S~J and Min B~I 2000 {\em Phys. Rev. B\/} {\bf 62}(1) 357--360
  \urlprefix\url{https://link.aps.org/doi/10.1103/PhysRevB.62.357}

\bibitem{CoS2exp}
Schr{\"o}ter N~B~M, Robredo I, Klemenz S, Kirby R~J, Krieger J~A, Pei D, Yu T,
  Stolz S, Schmitt T, Dudin P, Kim T~K, Cacho C, Schnyder A, Bergara A, Strocov
  V~N, de~Juan F, Vergniory M~G and Schoop L~M 2020 {\em Science Advances\/}
  {\bf 6} (\textit{Preprint}
  \eprint{https://advances.sciencemag.org/content/6/51/eabd5000.full.pdf})
  \urlprefix\url{https://advances.sciencemag.org/content/6/51/eabd5000}

\bibitem{Arc_inject_1}
Li P, Wu W, Wen Y, Zhang C, Zhang J, Zhang S, Yu Z, Yang S~A, Manchon A and
  Zhang X~x 2018 {\em Nature Communications\/} {\bf 9} 3990 ISSN 2041-1723
  \urlprefix\url{https://doi.org/10.1038/s41467-018-06518-1}

\bibitem{doi:10.1063/1.1656326}
Miyahara S and Teranishi T 1968 {\em Journal of Applied Physics\/} {\bf 39}
  896--897 (\textit{Preprint} \eprint{https://doi.org/10.1063/1.1656326})
  \urlprefix\url{https://doi.org/10.1063/1.1656326}

\bibitem{Teruya_2017}
Teruya A, Suzuki F, Aoki D, Honda F, Nakamura A, Nakashima M, Amako Y, Harima
  H, Uchima K, Hedo M, Nakama T and Yonuki 2017 {\em Journal of Physics:
  Conference Series\/} {\bf 807} 012001
  \urlprefix\url{https://doi.org/10.1088/1742-6596/807/1/012001}

\bibitem{BCS1}
Perez-Mato J, Orobengoa D, Tasci E, De~la Flor~Martin G and Kirov A 2011 {\em
  Bulgarian Chemical Communications\/} {\bf 43} 183--197

\bibitem{BCS2}
Aroyo M~I, Perez-Mato J~M, Capillas C, Kroumova E, Ivantchev S, Madariaga G,
  Kirov A and Wondratschek H 01 Jan 2006 {\em Zeitschrift für Kristallographie
  - Crystalline Materials\/} {\bf 221} 15 -- 27
  \urlprefix\url{https://www.degruyter.com/view/journals/zkri/221/1/article-p15.xml}

\bibitem{BCS3}
Aroyo M~I, Kirov A, Capillas C, Perez-Mato J~M and Wondratschek H 2006 {\em
  Acta Crystallographica Section A\/} {\bf 62} 115--128
  \urlprefix\url{https://doi.org/10.1107/S0108767305040286}

\bibitem{VASP1}
Kresse G and Hafner J 1993 {\em Phys. Rev. B\/} {\bf 47}(1) 558--561
  \urlprefix\url{https://link.aps.org/doi/10.1103/PhysRevB.47.558}

\bibitem{VASP2}
Kresse G and Hafner J 1994 {\em Phys. Rev. B\/} {\bf 49}(20) 14251--14269
  \urlprefix\url{https://link.aps.org/doi/10.1103/PhysRevB.49.14251}

\bibitem{VASP3}
Kresse G and Furthmüller J 1996 {\em Computational Materials Science\/} {\bf
  6} 15--50 ISSN 0927-0256
  \urlprefix\url{https://www.sciencedirect.com/science/article/pii/0927025696000080}

\bibitem{VASP4}
Kresse G and Furthm\"uller J 1996 {\em Phys. Rev. B\/} {\bf 54}(16)
  11169--11186
  \urlprefix\url{https://link.aps.org/doi/10.1103/PhysRevB.54.11169}

\bibitem{Brown_2005}
Brown P~J, Neumann K~U, Simon A, Ueno F and Ziebeck K~R~A 2005 {\em Journal of
  Physics: Condensed Matter\/} {\bf 17} 1583--1592
  \urlprefix\url{https://doi.org/10.1088/0953-8984/17/10/013}

\bibitem{Wannier90}
Pizzi G, Vitale V, Arita R, Blügel S, Freimuth F, G{\'{e}}ranton G, Gibertini
  M, Gresch D, Johnson C, Koretsune T, Iba{\~{n}}ez-Azpiroz J, Lee H, Lihm J~M,
  Marchand D, Marrazzo A, Mokrousov Y, Mustafa J~I, Nohara Y, Nomura Y,
  Paulatto L, Ponc{\'{e}} S, Ponweiser T, Qiao J, Thöle F, Tsirkin S~S,
  Wierzbowska M, Marzari N, Vanderbilt D, Souza I, Mostofi A~A and Yates J~R
  2020 {\em Journal of Physics: Condensed Matter\/} {\bf 32} 165902
  \urlprefix\url{https://doi.org/10.1088%2F1361-648x%2Fab51ff}

\bibitem{WannierTools}
Wu Q, Zhang S, Song H~F, Troyer M and Soluyanov A~A 2018 {\em Computer Physics
  Communications\/} {\bf 224} 405 -- 416 ISSN 0010-4655
  \urlprefix\url{http://www.sciencedirect.com/science/article/pii/S0010465517303442}

\bibitem{Naumov_origin}
Naumov I~I and Hemley R~J 2016 {\em Phys. Rev. Lett.\/} {\bf 117}(20) 206403
  \urlprefix\url{https://link.aps.org/doi/10.1103/PhysRevLett.117.206403}

\bibitem{Weyl_app_1}
Hills R~D~Y, Kusmartseva A and Kusmartsev F~V 2017 {\em Phys. Rev. B\/} {\bf
  95}(21) 214103
  \urlprefix\url{https://link.aps.org/doi/10.1103/PhysRevB.95.214103}

\bibitem{Weyl_app_2}
de~Juan F, Grushin A~G, Morimoto T and Moore J~E 2017 {\em Nature
  Communications\/} {\bf 8} 15995 ISSN 2041-1723
  \urlprefix\url{https://doi.org/10.1038/ncomms15995}

\bibitem{Weyl_app_4}
Rees D, Manna K, Lu B, Morimoto T, Borrmann H, Felser C, Moore J~E, Torchinsky
  D~H and Orenstein J 2020 {\em Science Advances\/} {\bf 6} eaba0509

\bibitem{GdPtBi}
Hirschberger M, Kushwaha S, Wang Z, Gibson Q, Liang S, Belvin C, Bernevig B~A,
  Cava R~J and Ong N~P 2016 {\em Nature Materials\/} {\bf 15} 1161--1165 ISSN
  1476-4660 \urlprefix\url{https://doi.org/10.1038/nmat4684}

\bibitem{Weyl_app_3}
Zhao B, Karpiak B, Khokhriakov D, Johansson A, Hoque A~M, Xu X, Jiang Y, Mertig
  I and Dash S~P 2020 {\em Advanced Materials\/} {\bf 32} 2000818
  (\textit{Preprint}
  \eprint{https://onlinelibrary.wiley.com/doi/pdf/10.1002/adma.202000818})
  \urlprefix\url{https://onlinelibrary.wiley.com/doi/abs/10.1002/adma.202000818}

\bibitem{Spin_arc_1}
Xu S~Y, Belopolski I, Sanchez D~S, Neupane M, Chang G, Yaji K, Yuan Z, Zhang C,
  Kuroda K, Bian G, Guo C, Lu H, Chang T~R, Alidoust N, Zheng H, Lee C~C, Huang
  S~M, Hsu C~H, Jeng H~T, Bansil A, Neupert T, Komori F, Kondo T, Shin S, Lin
  H, Jia S and Hasan M~Z 2016 {\em Phys. Rev. Lett.\/} {\bf 116}(9) 096801
  \urlprefix\url{https://link.aps.org/doi/10.1103/PhysRevLett.116.096801}

\bibitem{Spin_arc_2}
Huang Z, Lane C, Cao C, Zhi G~X, Liu Y, Matt C~E, Kuthanazhi B, Canfield P~C,
  Yarotski D, Taylor A~J and Zhu J~X 2020 {\em Phys. Rev. B\/} {\bf 102}(23)
  235167 \urlprefix\url{https://link.aps.org/doi/10.1103/PhysRevB.102.235167}

\end{thebibliography}

\end{document}